# Examining the effects of emotional valence and arousal on takeover performance in conditionally automated driving

Na Du

Industrial and Operations Engineering, University of Michigan

Feng Zhou

Industrial and Manufacturing Systems Engineering, University of Michigan-Dearborn

Elizabeth M. Pulver

State Farm Mutual Automobile Insurance Company

Dawn M. Tilbury

Mechanical Engineering, University of Michigan

Lionel P. Robert

School of Information, University of Michigan

Anuj K. Pradhan

Industrial and Mechanical Engineering, University of Massachusetts Amhest

X. Jessie Yang

Industrial and Operations Engineering, University of Michigan

**Manuscript type:** *Research Article*

**Running head:** *Emotions affect takeover performance*

**Word count:** 4500

**Corresponding author:** X. Jessie Yang, 1205 Beal Avenue, Ann Arbor, MI 48015, Email: xijyang@umich.edu

**Acknowledgment:** This work was supported by University of Michigan Mcity and in part by the National Science Foundation. The views expressed are those of the authors and do not reflect the official policy or position of State Farm[®].

# ABSTRACT

In conditionally automated driving, drivers have difficulty in takeover transitions as they become increasingly decoupled from the operational level of driving. Factors influencing takeover performance, such as takeover lead time and the engagement of non-driving related tasks, have been studied in the past. However, despite the important role emotions play in human-machine interaction and in manual driving, little is known about how emotions influence drivers' takeover performance. This study, therefore, examined the effects of emotional valence and arousal on drivers' takeover timeliness and quality in conditionally automated driving. We conducted a driving simulation experiment with 32 participants. Movie clips were played for emotion induction. Participants with different levels of emotional valence and arousal were required to take over control from automated driving, and their takeover time and quality were analyzed. Results indicate that positive valence led to better takeover quality in the form of a smaller maximum resulting acceleration and a smaller maximum resulting jerk. However, high arousal did not yield an advantage in takeover time. This study contributes to the literature by demonstrating how emotional valence and arousal affect takeover performance. The benefits of positive emotions carry over from manual driving to conditionally automated driving while the benefits of arousal do not.

# INTRODUCTION

According to the SAE standard (Society of Automotive Engineers, 2018), vehicles of Level 3 conditional automation and above are equipped with automated driving features. While people are still speculating if and when SAE Level 5 full automation will be ready (Sparrow & Howard, 2017), automated driving features at SAE Level 3, such as the Audi Traffic Jam Chauffeur, are expected to be introduced into the market (Bishop, 2019; Taylor, 2017).

With SAE Level 3 automation, drivers will no longer be required to actively monitor the driving environment and can engage in non-driving-related tasks (NDRTs). When the automated vehicle (AV) reaches its operational limits, however, drivers will have to take over control of the vehicle at a moment's notice. This transition of control represents the transfer of the longitudinal and lateral control responsibilities from the automated vehicle to the human driver, and usually involves the driver terminating NDRTs, moving eyes/hands/feet back to the road/steering wheel/pedals, and resuming control of the vehicle. Research indicates that drivers have difficulty in takeover transitions as they become increasingly decoupled from the operational level of driving (Ayoub, Zhou, Bao, & Yang, 2019; Eriksson & Stanton, 2017; Gold, Körber, Lechner, & Bengler, 2016; Petersen, Robert, Yang, & Tilbury, 2019; Zhou, Yang, & Zhang, 2019). In response to this known difficulty, research has been conducted to investigate factors affecting takeover performance, including the external driving environment (e.g., road elements, traffic situations, and weather conditions), types of NDRTs (e.g. reading, typing), individual characteristics (e.g., training, prior experience with automation, trust in automation, age), and design of human-machine interface (e.g., multi-modal display) (Eriksson & Stanton, 2017; Gold et al., 2016; Helldin, Falkman, Riveiro, & Davidsson, 2013; Körber, Gold, Lechner, & Bengler, 2016; Wan & Wu, 2018).

However, despite the important role emotion plays in human-machine interaction (Picard, 2003; Stickney, 2009) and in manual driving (Abdu, Shinar, & Meiran, 2012; Chan & Singhal, 2013; Jeon, 2017; Pêcher, Lemercier, & Cellier, 2009; Trick, Brandigampola, & Enns, 2012), little is known about how emotions influence drivers' takeover performance. The present study, therefore, aims to fill the research gap and examine the effect of emotional valence and arousal on takeover performance in conditionally automated driving.

**Emotion as a two-dimensional construct**

According to Russell (1980), emotion has at least two dimensions. The first dimension is valence, or how negative or positive a stimulus is. For example, watching a baby smiling is more positive than seeing a patient dying. The second dimension is arousal, or how sleepy or exciting a stimulus is. For example, listening to rock bands is associated with higher arousal than listening to meditation music. The two dimensions can be mapped in a two-dimensional space and the combinations of different values of valence and arousal are associated with different discrete emotions. For example, the upper left corner of the two-dimensional space represents emotions of negative valence and high arousal, such as anger; the lower right corner of the space represents emotions of positive valence and low arousal, such as calmness.

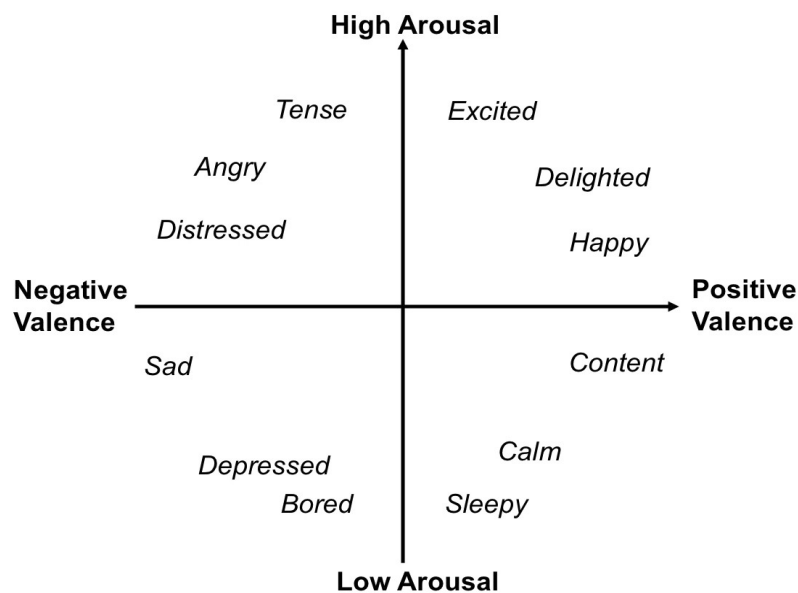

*Figure 1*. The circumplex model of emotion (Russell, 1980)

While it may be more straightforward for individuals to report discrete emotions they are experiencing, the dimensional view of emotion provides a more fundamental explanation of the relationships between emotions and behaviors (Barrett, 1998). The dimensional view is also supported by studies in neuroscience. With evidence of neural activities in the brain, valence and arousal appear to influence cognitive processes and human behaviors via distinct mechanisms (Dolcos, LaBar, & Cabeza, 2004; Kensinger, 2004). In line with this theoretical background, the current study manipulated and examined the effects of emotional valence and arousal systematically, instead of focusing on certain discrete emotions.

## Emotions in Manual and Automated Driving

The literature on how emotions affect human-machine interaction has grown significantly in the past two decades (Ahn, 2010; Picard, 2003; Stickney, 2009). One of the most important effects of emotions lies in its ability to capture attention. People tend to pay more attention to stimuli and thoughts that are more relevant to their current emotional states (Bower & Forgas, 2000). In addition, emotion has also been shown to influence memory (i.e., Emotional stimuli are generally remembered better than unemotional events (Reeves, Newhagen, Maibach, Basil, & Kurz, 1991; Thorson & Friestad, 1989)), cognitive style and performance (Rusting, 1998), and judgement and decision making (Peters, Västfjäll, Gärling, & Slovic, 2006).

Manual driving is a complex task involving attention, information processing, and action-based judgment. Drivers can become emotional on the road when they interact with external environments and other road users, which may lead to enormous consequences (Jeon, 2017). Some studies placed an emphasis on the effects of specific emotions on manual driving, in particular, the effects of anger. Anger, as one of the most commonly experienced emotions during driving, has received a substantial amount of research attention. A recent survey study by the AAA Foundation for Traffic Safety found that nearly 80 percent of drivers expressed anger, aggression or road rage at least once in the previous year, which are significant contributors to fatal crashes (AAA Foundation for Traffic Safety, 2016). An analysis of naturalistic driving data showed that drivers in elevated emotional states, including anger, sadness, crying, and/or emotional agitation have an increased risk of a crash by 9.8 times (Dingus et al., 2016).

Moreover, experimental studies indicate that anger leads to risky and aggressive behaviors, such as speeding and traffic rule violation (Abdu et al., 2012; Deffenbacher, Deffenbacher, Lynch, & Richards, 2003; Hu, Zhu, Gao, & Zheng, 2018; Jeon, Walker, & Yim, 2014; Underwood, Chapman, Wright, & Crundall, 1999). For example, Abdu et al. (2012) conducted a driving simulator study with 15 licensed drivers and found that angry drivers crossed more yellow traffic lights and tended to drive faster. Similarly, Jeon et al. (2014) found that anger led to a significantly lower perceived safety and degraded driving performance (i.e., larger deviations from the center line and more violations of traffic rules).

Moreover, researchers went beyond specific emotions and systematically explored the effects of positive/negative valence, and high/low arousal on manual driving performance (Chan & Singhal, 2013; Hancock, Hancock, & Janelle, 2012; Pêcher et al., 2009; Trick et al., 2012; Ünal, de Waard, Epstude, & Steg, 2013). Chan and Singhal (2013) investigated the effects of emotional valence on driving. In their study, participants were responsible for longitudinal and lateral control of the vehicle. At the same time, they were asked to view words of positive, negative and neutral valence on roadside billboards, and later to recall as many words as possible. Results revealed that drivers recalled more negative words than positive words, suggesting that negative stimuli distracted drivers' attention more severely. In another study, participants drove and viewed emotional images concurrently. Viewing positive images led to better lateral control but also slower speeds when compared to negative images (Hancock et al., 2012). The positive association between better vehicle control and positive valence was also reported in the studies of Trick et al. (2012) and Groeger (2013). Interestingly, using another emotion induction method, Pêcher et al. (2009) asked drivers to listen to music and found that happy music (positive valence) resulted in an unexpected large decrement of speed and a deteriorated lateral control in comparison with sad music (negative valence). The reason for the inconsistent findings could be due to the differences in participants' emotion induction methods (Steinhauser et al., 2018) and participants' base emotions and personal experience.

In addition, Trick et al. (2012) manipulated both emotional valence and arousal in an experiment where participants were exposed to a variety of images that were either positive or negative in valence and either high or low in arousal. After viewing the images, participants needed to brake in reaction to the sudden deceleration of a lead vehicle. Results showed that higher arousal led to faster hazard response if the hazard was presented shortly after viewing an image. Similarly, Navarro, Osiurak, and Reynaud (2018) and Ünal et al. (2013) conducted experiments to manipulate drivers' emotional arousal using musical tempo. Results showed that in a car following task, arousing musical background improved drivers' responsiveness to the speed changes of the followed vehicle compared to relaxing music.

Research in manual driving suggests the associations between positive valence and better vehicle control, and between higher arousal and faster hazard response. Despite the large amount of research on manual driving, few studies have examined how emotions influence driving performance in conditionally automated driving. In addition, those few studies are primarily focused on algorithm development to automatically detect drivers' emotional states by analyzing their physiological data. For example, Izquierdo-Reyes, Ramirez-Mendoza, Bustamante-Bello, Pons-Rovira, and Gonzalez-Vargas (2018) developed an algorithm using drivers' faces and electroencephalogram (EEG) data as features for classifier training. The results showed that a K Nearest Neighbors algorithm was able to recognize nine different emotions (neutral, anger, disgust, fear, joy, sadness, surprise, amusement, and anxiety) with an accuracy of approximately 97%.

Although emotion is likely to play an important role in takeover transitions relating to drivers' perception of the surrounding world, cognitive processing and decision making upon the takeover request, no studies have been conducted on this topic to our knowledge.

**The Present Study**

This study aimed to examine how emotions affect drivers' takeover performance in conditionally automated driving. With level 3 automation, drivers could potentially perform NDRTs when the automation mode is activated (Society of Automotive Engineers, 2018). However, they need to resume control of the vehicle within seconds if the vehicle reaches its performance limit. After receiving a takeover request (TOR), drivers need to quickly shift their attention to the road, process and comprehend the traffic situation, and select and execute an appropriate action. Given the drivers' tasks in takeover transitions, we have the following hypotheses.

We base our first hypothesis on the "broaden-and build" theory (Fredrickson & Branigan, 2005; Rowe, Hirsh, & Anderson, 2006), that positive emotions prompt individuals to broaden their focus of attention to the global aspects of an event and their thought-action repositories, whereas negative emotions narrow them. In takeover transitions, the broadened attention and thought-action repositories aid drivers in traffic situation comprehension and action selection, and hence enhance the takeover quality. Therefore, we hypothesize:

**H1:** positive emotions will enhance takeover quality in terms of driving smoothness, ride comfort, and collision risk.

Research in manual driving showed that high arousal led to faster response time in hazard detection (Navarro et al., 2018; Trick et al., 2012; Ünal et al., 2013). Notably in manual driving, drivers allocated and managed their attention between the driving task and the NDRTs without any alerts or alarms. In conditionally automated driving, upon receiving a takeover request (TOR), drivers are required to immediately switch their attention from the NDRTs to the driving task, and drivers can respond to a TOR reflexively (Zeeb, Buchner, & Schrauf, 2016). Thus, we hypothesize:

**H2:** The advantage of high arousal in faster response time will be reduced or even diminished in takeover transitions.

To test the hypotheses, we conducted a human-subject experiment with 32 drivers using a fixed-based driving simulator. Participants drove a vehicle with Level 3 automation and watched movie clips for emotion induction. Each of them experienced four takeover events and their takeover time and quality were recorded and analyzed.

## METHOD

This research complied with the American Psychological Association code of ethics and was approved by the Institutional Review Board of the University of Michigan.

### Participants

A total of 32 university students (average age = 21.4 years, SD = 2.9; 17 females and 15 males) with normal or corrected-to-normal vision participated in the experiment. Participants were screened for valid US driver license status and susceptibility to simulator sickness. The study lasted 60 to 75 minutes, and each participant was compensated with $30 upon completion of the experiment.

### Apparatus and Stimuli

The study was conducted using a fixed-based desktop driving simulator. The simulator ran the SimCreator driving simulation engine from Realtime Technologies Inc. (RTI, Michigan, USA) (Figure 2). To present the driving environment to participants, forward road scenes were displayed on a 32-inch computer monitor about 2.5 feet in front of the driver. A rear-view image was also displayed in a separate window on the forward screen. The simulated vehicle was controlled by a Logitech steering wheel and pedal system connected via USB interface to the SimCreator components.

In this study, the automation features of the driving simulator were programmed to simulate an SAE Level 3 AV, wherein the AV performed the longitudinal and lateral vehicle control, navigated, and responded to traffic control devices and other traffic elements, and the driver was not required to actively monitor the driving environment. However, there were unexpected events that the AV could not handle and would request the driver to take over control of the vehicle.

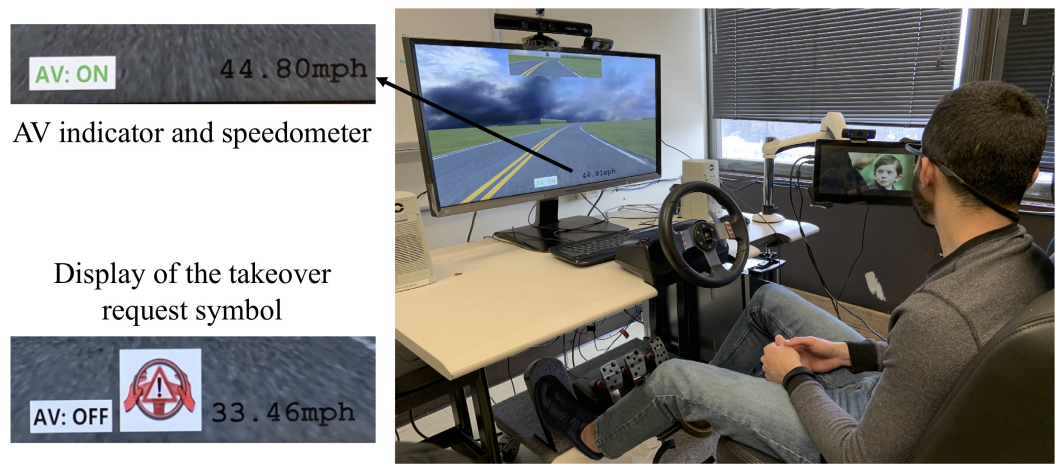

*Figure 2.* Illustration of driving simulator in the experiment

**Experimental Design**

The experiment used a within-subjects design in order to minimize effects of extraneous variables and to increase statistical power. We induced different values of emotional valence and arousal within each participant, covering the four quadrants of the valence-arousal space (Figure 1).

As shown in Table 1, eight 4-minute movie clips were selected for emotion induction based on prior literature (Ekman, Freisen, & Ancoli, 1980; Gross & Levenson, 1997; Lisetti & Nasoz, 2004; Uhrig et al., 2016). To minimize the ordering effect, the sequence of the emotion induction conditions was counterbalanced using a Latin square design. Among all the participants, 10 participants had previously watched 1 out of the 8 movies and 5 participants watched 2 out of the 8 movies before the study. Based on participants' comments in the post-experiment debriefing sessions, they were actively engaged in watching the movie clips.

Drivers were ensured that there was no need to monitor the driving environment when the AV was in the automation mode, and they could focus on watching the videos until a TOR was issued. The TOR was in both auditory and visual format: an audible spoken phrase "takeover" and an icon representing a pair of red hands on a red steering wheel (Kuehn, Vogelpohl, & Vollrath, 2017) (Figure 2). The sound volume and visual intensity were the same for all the participants throughout the experiment. Each participant confirmed in the practice drives that the TOR could be heard clearly while the participant was watching the movie clips. The speedometer, the AV mode indicator and the TOR symbol were displayed in real time at the lower center of the screen.

TABLE 1: *Descriptions of movie clips used for emotion induction*

| Emotion | Movie | Scene | Citation |
|---------|-------|-------|----------|
| Sad | The Champ | Death of the Champ | Lisetti and Nasoz (2004) |
| | Finding Neverland | Death of the boy's mother | Uhrig et al. (2016) |
| Angry | Schindler's list | Woman engineer being shot | Lisetti and Nasoz (2004) |
| | Cry Freedom | Innocent people being shot | Gross and Levenson (1997) |
| Happy | Bruce Almighty | Man getting power from the God | Uhrig et al. (2016) |
| | Big Fish | Boy expressing love to the girl | Uhrig et al. (2016) |
| Calm | ScreenPeace screensaver with city scenes | | Gross and Levenson (1997) |
| | Beautiful trees and flowers in the world | | Ekman et al. (1980) |

Each participant went through four takeover events. The takeover events were designed based on prior literature (Koo, Shin, Steinert, & Leifer, 2016; Miller et al., 2016; Rezvani et al., 2016) (Table 2). In the AV mode, the vehicle always drove in the right lane. The TOR was issued when the AV was 4 seconds away from the construction zone/bicyclist/police vehicle/swerving vehicle. Participants were expected to change to the left lane during the takeover transitions.

TABLE 2: *Descriptions of takeover events*

| Environment | Event descriptions |
|-------------|-------------------|
| Urban | Construction zone ahead |
| Urban | Bicyclist in the lane ahead |
| Rural | Police vehicle on shoulder |
| Rural | Swerving vehicle ahead |

**Measures**

We measured participants' subjective ratings of emotional valence and arousal, as well as objective measures of their takeover performance.

The Self-Assessment Manikin (SAM) instrument was used to assess participants' emotional valence (1 = extremely negative, 9 = extremely positive) and arousal (1 =

lowest arousal, 9 = highest arousal) based on their emotions induced by the movie clips (Bradley & Lang, 1994).

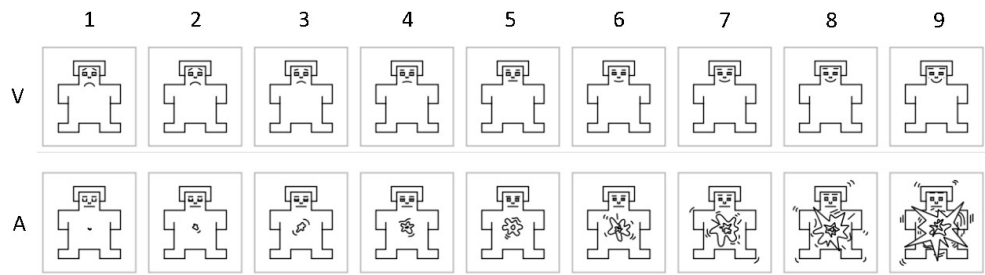

*Figure 3*. The Self-Assessment Manikin (SAM) (Bradley & Lang, 1994)

Takeover performance was assessed in the timing and quality aspects. Takeover time was calculated as the time between the TOR and the start of the maneuver. The start of the maneuver is defined as a 2-degree change in steering wheel angle or a 10% depress of pedals, whichever is first (Gold et al., 2016). Takeover quality was assessed by three representative driving measures in the obstacles ahead scenarios: maximum resulting acceleration, maximum resulting jerk, minimum time to collision (TTC$_{min}$) within the time window between the TOR and the end of the lane changing behavior (i.e. the center of the vehicle reached the boundary of the other lane). Consistent with prior research (Hergeth, Lorenz, & Krems, 2017), maximum resulting acceleration is calculated as

$max\ acceleration_{resulting} = \max_{t}\sqrt{acceleration^2_{longitudinal} + acceleration^2_{lateral}}$. A smaller acceleration represents a smoother and safer reaction to TORs. In addition, we calculated the maximum resulting jerk as

$max\ jerk_{resulting} = \max_{t}\sqrt{jerk^2_{longitudinal} + jerk^2_{lateral}}$. Jerk is the derivative of acceleration and has been utilized to evaluate shift quality, ride comfort (Huang & Wang, 2004) and driving aggressiveness (Bagdadi & Várhelyi, 2011, 2013; Feng et al., 2017). Similarly, a smaller jerk represents higher takeover quality. TTC is a time-based safety indicator for detecting rear-end collision risk and is defined as the time taken for two objects to collide if maintaining their present speeds and trajectories (Hayward, 1972).

Five crashes occurred in the study. Under such situations, minimum TTC was treated as "not applicable", and other driving dynamic variables were calculated using the time window between the TOR and the time when drivers re-engaged the automation mode.

**Experimental procedure**

Once participants arrived at the lab, they signed an informed consent and filled a demographic form. Next, participants received a 5-minute training, where they practiced how to change lanes and engage/disengage the automated driving mode via pressing a button on the steering wheel. They were asked to comply with all the traffic laws when they drove manually and the speed limit was 35 mph. Participants also experienced an unexpected takeover event in the practice while watching a 1-minute movie clip of Zen Garden. The movie clip was played on a tablet located on the right side of the driver's seat. The takeover event was the scenario where the traffic lights at an intersection did not work, and required the driver to observe the surroundings and drive manually. Participants were asked to re-engage the AV once they had negotiated the drive.

After the training session, participants completed two drive courses, each containing two takeover events. As shown in Figure 4, each course began with the command to activate the automated driving mode. Then there was an emotion induction phase when participants were asked to watch two 4-minute movie clips aimed at inducing the same emotion. Close to the end of the movie clips, a takeover request was issued, and participants were required to take over control of the vehicle immediately. Once participants negotiated the drive, they could hand over the control back to the AV. After participant re-engaged the AV mode, they were asked to recall the scenes in the movie clips and complete the SAM survey to indicate their emotional valence and arousal when watching the movie clips.

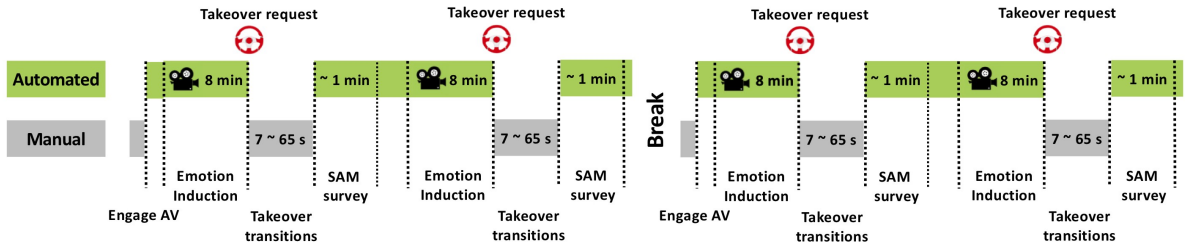

*Figure 4*. Sequence of takeover events in the experiment

## RESULTS

Data from one participant was excluded from analysis as the participant did not follow the instructions from the experimenter. All hypotheses were tested using data from the remaining 31 participants. The SAM Likert scales from 1 to 9 (low arousal: 1-4, high arousal: 6-9; negative valence: 1-4; positive valence: 6-9) were normalized to a scale from -1 to 1 (Miranda Correa, Abadi, Sebe, & Patras, 2018). Data points with 0 valence or 0 arousal were deleted. Figure 5 shows the distribution of valence and arousal values in the four quadrants: positive valence high arousal, negative valence high arousal, negative valence low arousal, and positive valence low arousal, respectively.

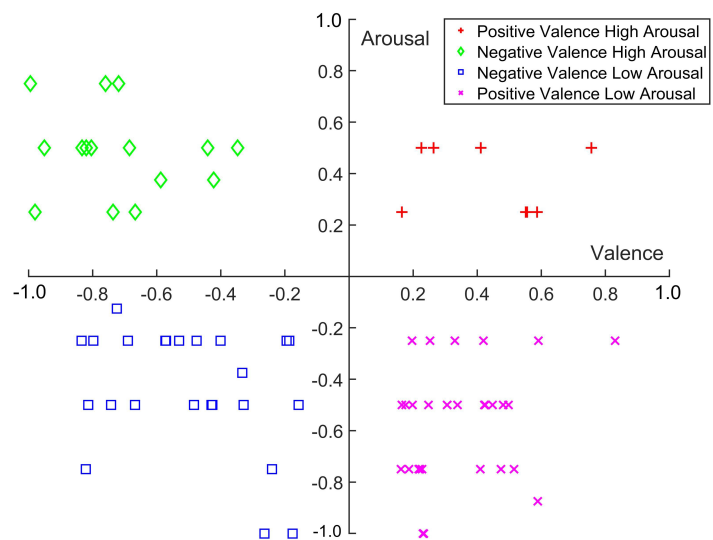

*Figure 5*. Raw data of subjective ratings in the valence-arousal plane

We used a mixed linear model to analyze the relationship between valence, arousal and takeover performance (timeliness and quality). Results are reported as significant for $\alpha < .05$. Table 3 summarizes the mean and standard error (SE) values of the dependent measures.

TABLE 3: *Mean and Standard Error (SE) values of dependent measures*

| | Negative Valence | | Positive Valance | |
| --- | --- | --- | --- | --- |
| | Low arousal | High arousal | Low arousal | High arousal |
| Takeover time (s) | $1.88 \pm .09$ | $1.69 \pm .07$ | $1.79 \pm .07$ | $1.88 \pm .16$ |
| Max resulting acceleration (m/$s^2$) | $6.56 \pm .75$ | $6.14 \pm .81$ | $5.85 \pm .62$ | $3.95 \pm .79$ |
| Max resulting jerk (m/$s^3$) | $113 \pm 18$ | $115 \pm 21$ | $94 \pm 17$ | $42 \pm 16$ |
| TTC$_{min}$ (s) | $.98 \pm .15$ | $.77 \pm .15$ | $.72 \pm .12$ | $.67 \pm .15$ |

**Takeover time**. No significant effect was found for either valence ($F(1, 57) = .04, p = .84$) or arousal ($F(1, 76) = .32, p = .57$), and the interaction effect was not significant ($F(1, 63) = .47, p = .50$).

**Takeover quality**. With regard to the maximum resulting acceleration, there was a significant effect of valence ($F(1, 56) = 4.26, p = .04$). Positive valence led to a smaller maximum resulting acceleration (Figure 6). In addition, there was a trend that high arousal led to a smaller maximum resulting acceleration ($F(1, 77) = 3.24, p = .08$). The interaction effect of valence and arousal on maximum resulting acceleration was not significant ($F(1, 64) = .79, p = .38$).

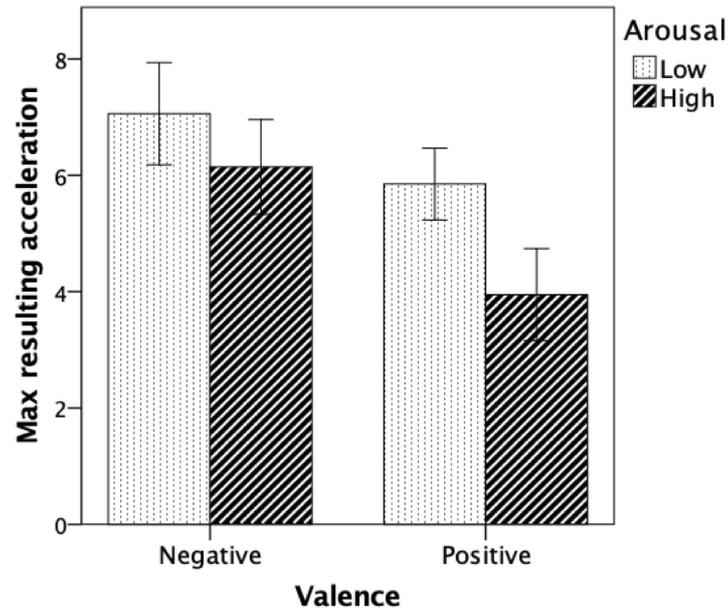

*Figure 6*. Maximum resulting acceleration (m/$s^2$)

There was a significant effect of valence on maximum resulting jerk ($F$(1, 55) = 6.47, $p$ = .01), with positive valence leading to a smaller maximum resulting jerk (Figure 7). The main effect of arousal ($F$(1, 73) = 1.84, $p$ = .18) and the interaction effect were non-significant ($F$(1, 61) = 1.71, $p$ = .20).

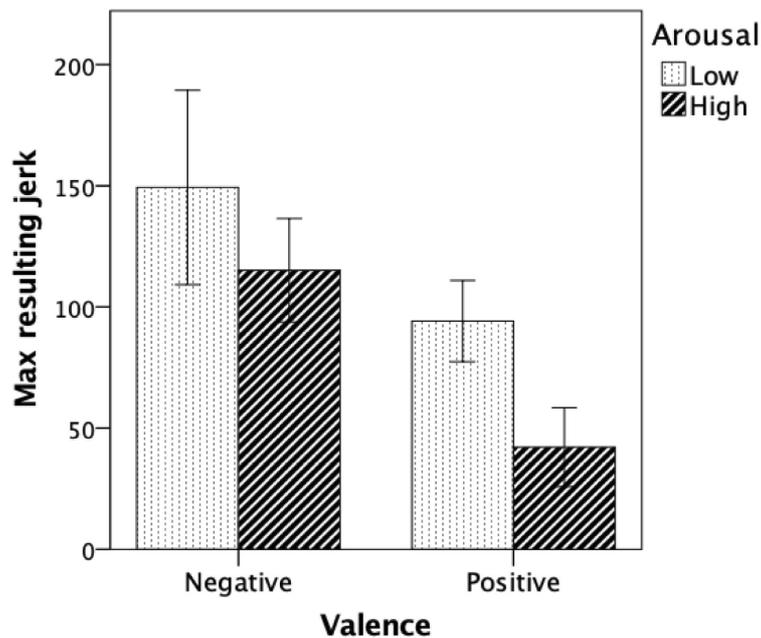

*Figure 7*. Maximum resulting acceleration (m/$s^3$)

There were no significant effects of valence ($F$(1, 57) = 1.19, $p$ = .28) and arousal

$(F(1, 75) = .67, p = .42)$ on $TTC_{min}$. The interaction effect was not significant $(F(1, 65) = .31, p = .58)$.

## DISCUSSION

Drivers' tasks in manual driving and conditionally automated driving (Level 3 AV) are fundamentally different. In manual driving, drivers continuously perform lateral and longitudinal control. Therefore, prior studies in manual driving mainly treated the NDRTs as distractions. With conditionally automated driving, however, drivers largely perform a single task (i.e. either the driving task or the NDRT) at any one time. When the automation mode is on, drivers can perform any NDRT at their own discretion. After receiving a TOR, drivers are expected to relinquish the NDRT and resume the driving task immediately. The distinction between manual driving and conditionally automated driving suggests findings in manual driving cannot be directly applied to takeover transitions. In the present study, we hypothesized that positive valence will lead to better takeover quality but the benefit of high arousal on response time would be reduced.

Our first hypothesis was built on the "broad-and-build" theory (Fredrickson & Branigan, 2005; Rowe et al., 2006) stating that positive emotions prompt individuals to broaden their focus of attention and their thought-action repositories. The broader span of attention enables drivers to perceive and process different stimuli in the traffic situation and avoid tunnel vision. Larger thought-action repositories allow drivers to identify a more appropriate action given a specific traffic situation. Our findings largely support the first hypothesis, that positive valence led to better takeover performance, reflected by a smaller maximum resulting acceleration and a smaller maximum resulting jerk. Smaller maximum resulting acceleration and maximum resulting jerk are associated with a higher level of safety (Hergeth et al., 2017), shift quality, and ride comfort (Huang & Wang, 2004), and lower driving aggressiveness (Bagdadi & Várhelyi, 2011, 2013; Feng et al., 2017). The results are in line with some studies examining the effects of valence in manual driving, where positive valence led to better vehicle control (Chan & Singhal, 2013; Groeger, 2013; Hancock et al., 2012; Trick et al., 2012), suggesting that the benefits of positive valence can be carried over from manual driving to automated driving. However, we failed to find any difference in $TTC_{min}$, and the reason could be explained as follows. Time to collision represents the time taken for two objects to collide with each other and is an indicator of collision risk. With a negative emotion, drivers' attentional focus and thought-action repositories are narrowed. Therefore, they might employ immediate survival-oriented behaviors and brake abruptly, leading to a potentially larger $TTC_{min}$.

In the present study, we adopted three metrics aimed to assess driving smoothness, ride comfort and collision risk. However, we notice the wide range of metrics used to measure takeover quality in prior literature, including crash rates, different statistics of velocity, acceleration, jerk, and TTC (Please refer to McDonald et al. (2019) for the detailed list). This wide range of metrics makes it difficult to summarize findings in prior literature. There is an urgent need to examine if possible and how to propose a standard sets of metrics for measuring takeover quality.

We also hypothesized that the advantage of high arousal in response time should be reduced or even diminished in takeover transitions, which is supported by the non-significant effect of arousal on takeover time. Prior research in manual driving showed that high arousal led to faster response time in hazard detection (Navarro et al., 2018; Trick et al., 2012; Ünal et al., 2013). In takeover transitions, however, TORs serve as an attention management tool. Upon receiving a TOR, drivers are required to attend to the driving task. Moreover, in our study drivers were engaged in a hands-free task (i.e. watching movies) before the TOR was issued. Without the need to physically end the task and put down the NDRT device, drivers could immediately switch their attention from the tablet to the driving scene. Our results showed that this process took less than 2 seconds, no matter in which emotional state drivers were. Recent studies comparing different types of NDRTs on takeover quality and timeliness showed that the types of NDRTs only influenced the takeover quality and not takeover time (Bueno et al., 2016; Gold et al., 2016; Körber et al., 2016; Zeeb et al., 2016; Zeeb,

Härtel, Buchner, & Schrauf, 2017), providing further support for our findings.

To our surprise, the results suggest a trend that high arousal led to a smaller maximum resulting acceleration and thus better takeover quality. This implies that the benefits of high arousal in mobilizing attentional resources and effort for immediate actions could be reflected in takeover quality. Further research is needed to elucidate this effect.

## CONCLUSION

Drivers have difficulty in takeover transitions as they become increasingly decoupled from the operational level of driving. In response to this challenge, researchers have started to look into factors that could influence drivers' takeover performance. Despite the important role emotion plays in human-machine interaction and in manual driving, little is known about how emotion affects takeover performance in conditionally automated driving. By systematically manipulating drivers' emotional states, the current study extended earlier research by demonstrating how valence and arousal influence takeover time and quality. The benefits of positive emotions carry over from manual driving to conditionally automated driving while the benefits of arousal do not. Moreover, our study provides empirical evidence that with regard to emotions we cannot simply apply the findings in manual driving to automated driving.

Our findings have implications on the design of in-vehicle emotion regulation systems. Advances in machine learning enable accurate detection of drivers' emotional states in real-time (Izquierdo-Reyes et al., 2018; Picard, 2003). For example, if the system detects that a driver is a negative emotional state, strategies such as reappraisal and distraction (Naragon-Gainey, McMahon, & Chacko, 2017) can be used to help the driver manage his or her negative emotion, resulting in better takeover performance. Also, the AV could even delay or avoid handing over control.

The present study has several limitations that should be taken into consideration. First, we only examined one type of takeover event wherein the drivers perceived certain hazards and were expected to change lanes. Further research could be extended to other types of takeover events such as lane markings disappearing. Second, the study was conducted in a fixed-based desktop driving simulator with limited fidelity. Future studies can investigate the effects of emotional valence and arousal in a higher fidelity laboratory environment or a naturalistic driving environment. Third, drivers' emotional valence and arousal values were queried after a takeover event occurred. Although we followed a standard practice and asked the drivers to recall the movie clips, and based on which to indicate their emotional states prior to the takeover event, experiencing the event per se might influence drivers' perceptions of the movie clips. Further research could employ physiological measures of emotion, which could indicate drivers' emotional states non-intrusively before a takeover event. Meanwhile, eye-tracking metrics such as gaze behaviors could be recorded and analyzed in order to better understand drivers' attention allocation during conditionally automated driving.